\documentclass{I10479_elsart}

\usepackage{longtable}

\usepackage{natbib}

\newcounter{time}
\usepackage{graphicx}

\newcommand\fdg{\mbox{$.\!\!^\circ$}}%
\newcommand\farcs{ \mbox{%
  \kern  0.13ex.%
  \kern -0.95ex\raisebox{.9ex}{\scriptsize$\prime\prime$}%
  \kern -0.1ex%
 }%
}

\begin{document}

\begin{frontmatter}



\title{Tentative Detection of the Rotation of Eris}


\author[roe]{Henry G.\ Roe}, 
\author[pike]{Rosemary E.\ Pike},
\author[brown]{Michael E.\ Brown}

\address[roe]{Lowell Observatory,
Flagstaff, AZ 86001 (U.S.A.)}
\address[pike]{Gemini Observatory,
Hilo, HI  96720 (U.S.A.)}
\address[brown]{California Institute of Technology, 
Division of Geological and Planetary Sciences, Pasadena, CA 91125 (U.S.A.)}



 \begin{abstract}
  We report a multi-week sequence of B-band photometric measurements
  of the dwarf planet Eris using the {\it Swift} satellite.  The use
  of an observatory in low-Earth orbit provides better temporal
  sampling than is available with a ground-based telescope.  We find
  no compelling evidence for an unusually slow rotation period of
  multiple days, as has been suggested previously.  A $\sim$1.08 day
  rotation period is marginally detected at a modest level of
  statistical confidence ($\sim$97\%).  Analysis of the combination of
  the $Swift$ data with the ground-based B-band measurements of
  \citet{2007AJ....133...26R} returns the same period ($\sim$1.08~day)
  at a slightly higher statistical confidence ($\sim$99\%).

 \end{abstract}
 \begin{keyword}
KUIPER BELT
 \end{keyword}


\end{frontmatter}

\section{Introduction}

The recently discovered Kuiper belt object Eris is
1.05$\pm$0.4\% the size of Pluto \citep{2006ApJ...643L..61B}.  The
near-infrared spectrum of Eris is dominated by methane
\citep{2005ApJ...635L..97B}, suggesting that its surface, like Pluto,
is covered in significant deposits of methane frost.  The surface of
Pluto is variegated, with regions of low and high
albedo  \citep{cruikshank_plutobook}. 
 The heterogeneity of
Pluto's surface is revealed in its light curve, which has a large
amplitude of 0.33 magnitude \citep{1997Icar..125..233B}.  The 
V band geometric albedo of Eris (85$\pm$7\%; \citet{2006ApJ...643L..61B})
is approximately equal to the albedo of the brightest patches on
Pluto's surface \citep{1997AJ....113..827S}.  This led to the suggestion
that the surface of Eris is likely homogeneous and of a composition
similar to the brightest patches of Pluto \citep{2006ApJ...643L..61B}.

Pluto's rotation period (6.4 days) is set by tidal interactions with
its large moon Charon.  Dysnomia, the one known moon of Eris, is far
too small to have significantly altered the rotation rate of Eris.
Several previous observers have not conclusively 
identified the rotation period of Eris
 \citep{2006A&A...460L..39C,
  2007AdSpR..40..238L, 
  2007AJ....134..787S,2007AJ....133...26R,2008A&A...479..877D}.  
The goal of the
observations reported here was to measure the
rotation period of Eris.

\section{Observations}

In December 2006 and January 2007 we acquired a sequence of images
with the Ultraviolet/Optical Telescope (UVOT) of the {\it Swift}
spacecraft.  The {\it Swift} mission was designed to detect Gamma Ray
Bursts (GRB) and rapidly slew to measure their optical afterglow.
Accepting the risk that observations may be interrupted to follow an
evolving GRB, UVOT is available for non-GRB science.  While UVOT is
small (30 cm) compared with many groundbased telescopes, UVOT has two
distinct advantages for this work.  Being space-based UVOT is not
subject to the vagaries of weather or atmospheric transparency and is
therefore photometrically much more stable than a ground-based
telescope.  Additionally, being in low-Earth orbit UVOT can observe
throughout each 24 hour day, losing only $\sim$45 minutes out of every
$\sim$90 minute orbit, while a ground-based telescope is limited by
its site to observing Eris for 6-10 hours per day.  These are
particularly useful advantages when an object has a slow rotation
period of one day or longer, such as has been suggested for Eris.
Over several weeks we acquired nearly 200 ksec of exposure time on
Eris with UVOT in the B filter.  Most of these images were acquired in
a 2$\times$2 binning mode with a pixel size of 1$\farcs$0.  The first
three images of Table 1 were acquired with no binning
(0$\farcs$5/pixel) and were rebinned to 2$\times$2 for the analysis.

Following the photometric prescription of \citet{2006PASP..118...37L}
we measured the magnitude of Eris in each individual frame.  We found
220 frames taken between 19 December 2006 and 16 January 2007 were of
usable quality.  (See Table 1 for a full listing of the observations.)
To refine the precision of the frame-to-frame relative photometry we
also measured an ensemble of 26 comparison stars, chosen to be between
1 magnitude fainter and 2 magnitudes brighter than Eris and to appear
in a minimum of 180 of the 220 frames.  Several other stars that met
these criteria were eliminated for having obvious photometric
periodicities or trends.  None of these eliminated stars displayed
periodicities near the $\sim$1.1 day period we find for Eris.  The
mean full-width at half-maximum (FWHM) of Eris and the 26 comparison
stars was 2$\farcs$3.  We found the aperture radius for optimum
signal-to-noise ratio to be close enough to the 3$\farcs$0 aperture
radius recommended by \citet{2006PASP..118...37L} that we adopted a
3$\farcs$0 aperture radius throughout.  We determined the magnitude
correction for each frame using the ensemble photometry algorithm of
\citet{1992PASP..104..435H}.  With the photometric stability of UVOT
these corrections are small, but non-negligible.

 The Sun-Eris-Earth distance varied  over the time period of
observations and the reported measurements have been scaled to remove
this known effect ($<$1$\%$ over the time period).  The Sun-Eris-Earth
phase angle also varied  over the time period of observations
between 0$\fdg$535 and 0$\fdg$582 (see Table~\ref{swiftmaintable}).
The phase coefficient for Eris in the B-band is zero 
within uncertainty 
\citep[-0.004$\pm$0.028 mag deg$^{-1}$;][]{2007AJ....133...26R},
implying $\leq$0.0015 mag of variation in the {\it Swift} data due to 
phase angle variation.
Any brightness variation due to phase angle is well below our detection
limits, and we have not attempted to scale the data to remove the 
phase angle effect.

\begin{figure}[p!]
\begin{center}
\includegraphics[width=4in]{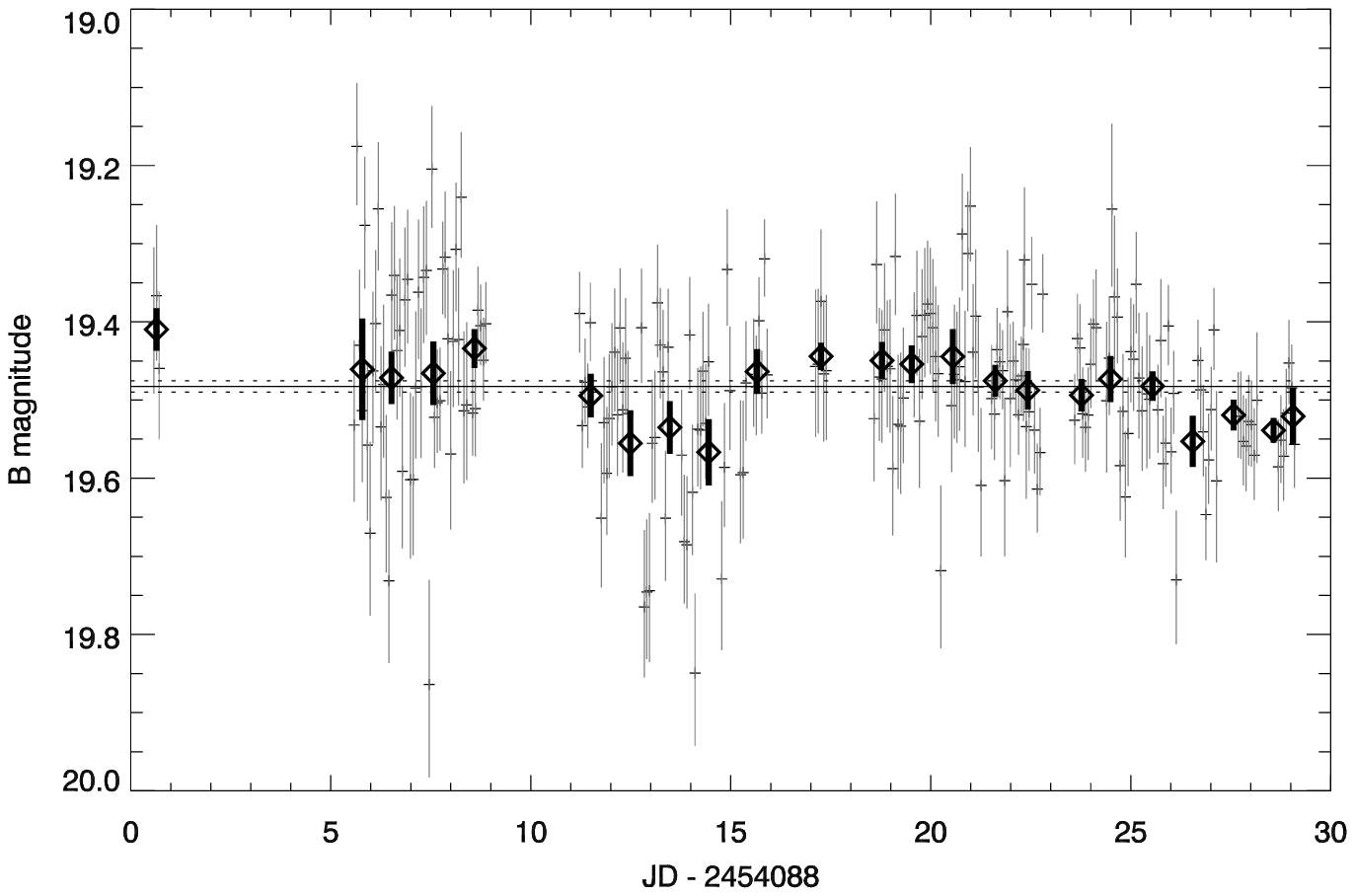}
\caption[Swift fluxes]{
	\label{swift_fluxes}
	\label{lastfig}			
	Each of the 220 {\it Swift} measurements of Eris in the $B$ filter
is shown along with its estimated photometric uncertainty. 
Overplotted as diamonds
 are the daily means of the data with uncertainties estimated from
the standard deviation within each day.
Also shown is the mean magnitude of the entire dataset, with 1$\sigma$ 
estimated uncertainty.
	}
\end{center}
\end{figure}

The measured count rate of photons from Eris is low,
approximately 0.6 photons sec$^{-1}$, and the estimated uncertainty in
any individual measurement is large, ranging between 0.05 and 0.13 mag with a 
median and mean of 0.08 mag.  This spread of uncertainty per measurement is
almost entirely explained by variable exposure times, necessitated by 
telescope and instrument scheduling issues.
The resulting
measurements of Eris are given in 
 Table~\ref{swiftmaintable} and shown in Fig.~\ref{swift_fluxes}.
Also shown in Fig.~\ref{swift_fluxes} are the mean daily measured
magnitudes of Eris.
Weighting the individual measurements by their estimated uncertainties
we find a mean B magnitude of Eris of 19.494$\pm$0.007, where the final
uncertainty is estimated from the standard deviation of the measurements
divided by $\sqrt{220}$.

\begin{figure}[p!]
\begin{center}
\includegraphics[width=4in]{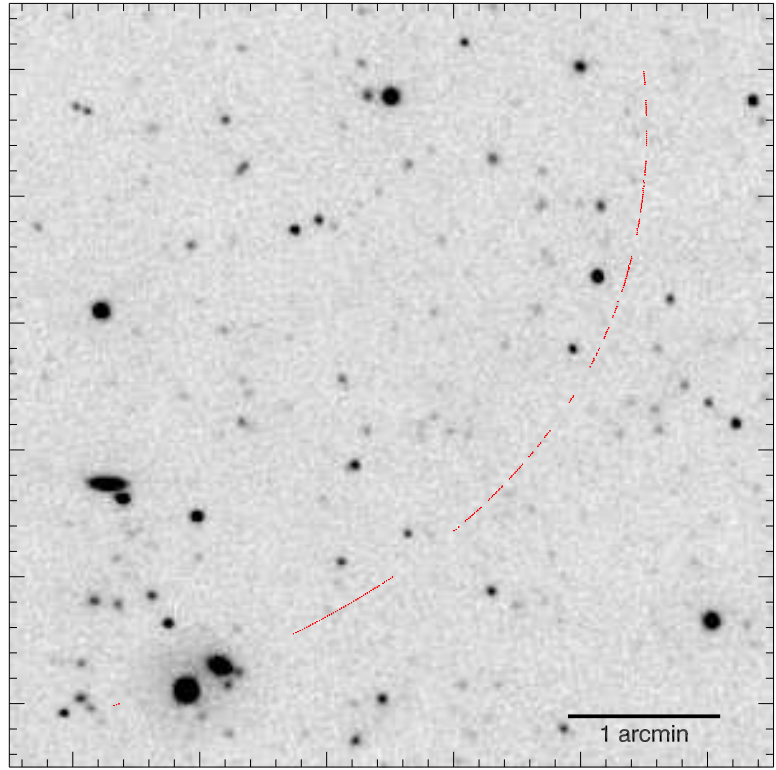}
\caption[Swift combined image]{
	\label{swift_image}
	\label{lastfig}			
Combination of all $Swift$ $B$ images.
The gray scale of the image is scaled linearly between $\pm$0.3\%
of the mean flux of Eris. Overplotted is the position of Eris
at the time of each exposure used in the analysis.
	}
\end{center}
\end{figure}

Throughout the observations Eris moved across the background field
of stars and galaxies and one possible explanation for any observed variability
 would
be the blending of background objects into the aperture used to measure
the flux of Eris.   Figure~\ref{swift_image} is a combined image of
the entire dataset with the positions of Eris during our observations
overplotted.  To test for possible
background source contamination we performed aperture photometry,
using the same technique as above, for each position on the combined
image where we had observed Eris.  The standard deviation of this
photometry was 0.0004~photons sec$^{-1}$ with a maximum excursion from
the mean of 0.0006~photons sec$^{-1}$.  Given a mean count rate of
$\sim$0.6~photons sec$^{-1}$ for Eris, the maximum background source
contribution to any observed variation in brightness on Eris is 
0.001 mag.  This is much smaller than the measured photometric uncertainties
and demonstrates that background stars and galaxies did not 
significantly contaminate the measured Eris fluxes.
Dysnomia, the known
moon of Eris, is far too faint to have influenced our measurements.

\section{Discussion}

\begin{figure}[p!]
\begin{center}
\includegraphics[width=3in]{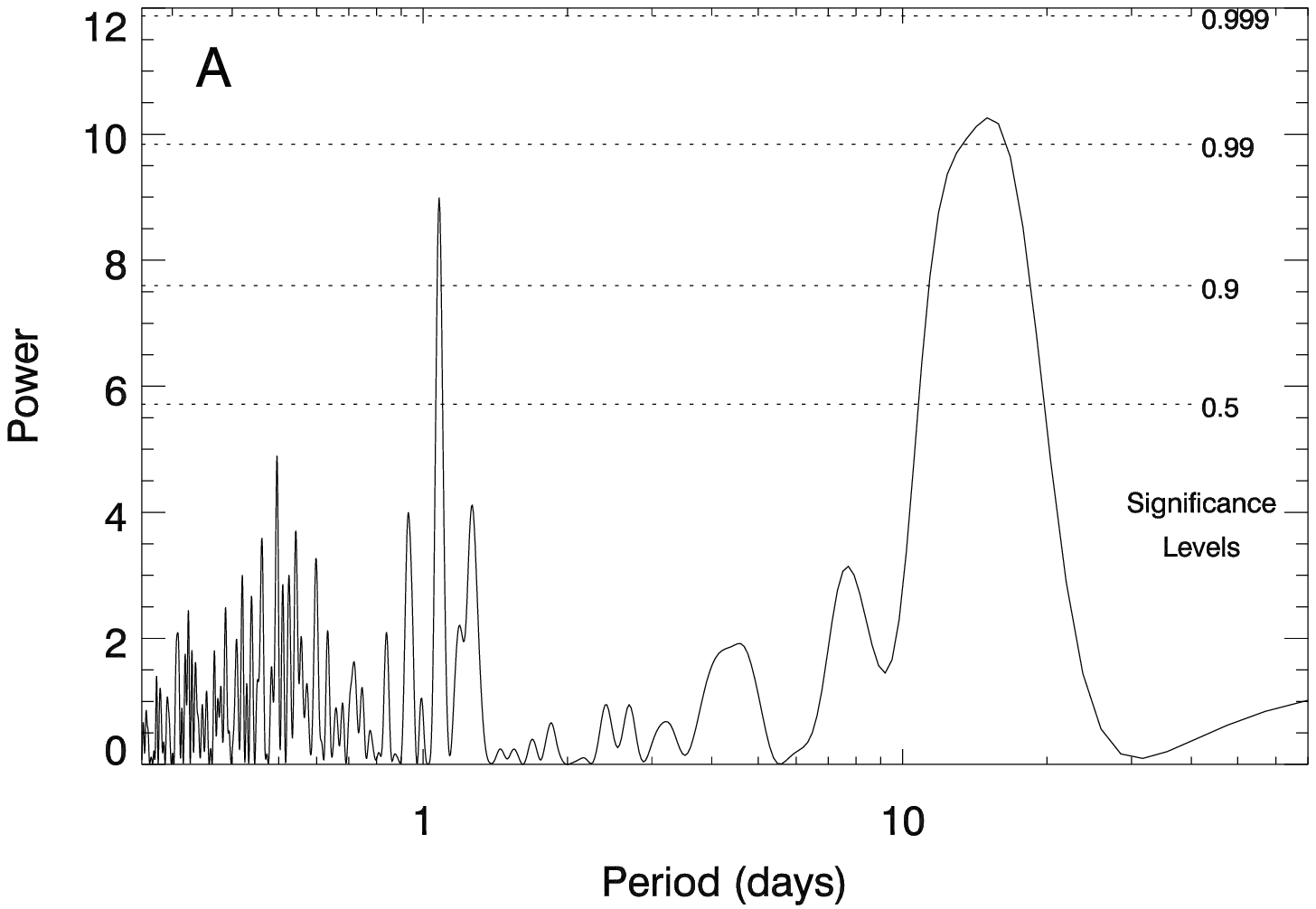}
\includegraphics[width=3in]{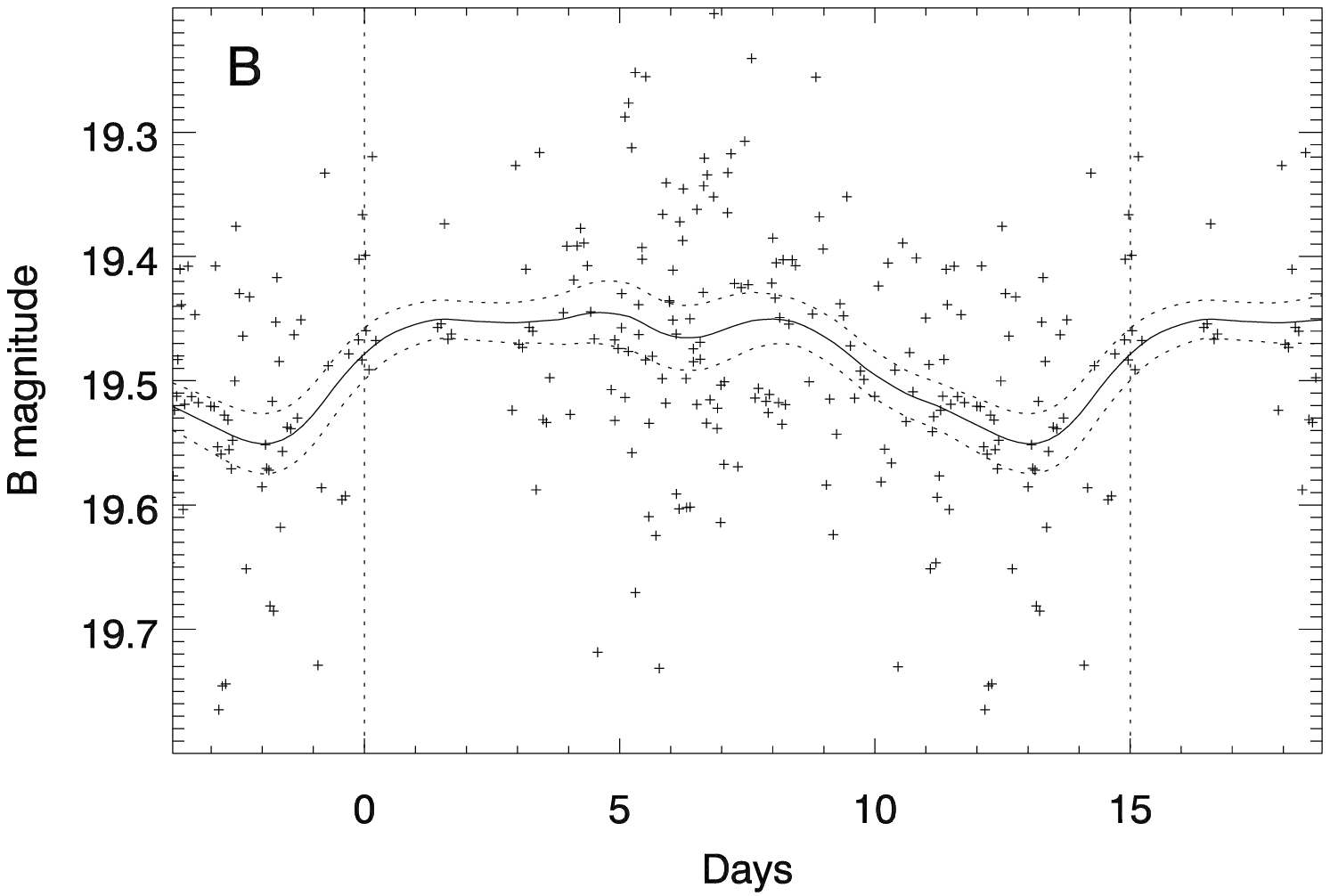}
\includegraphics[width=3in]{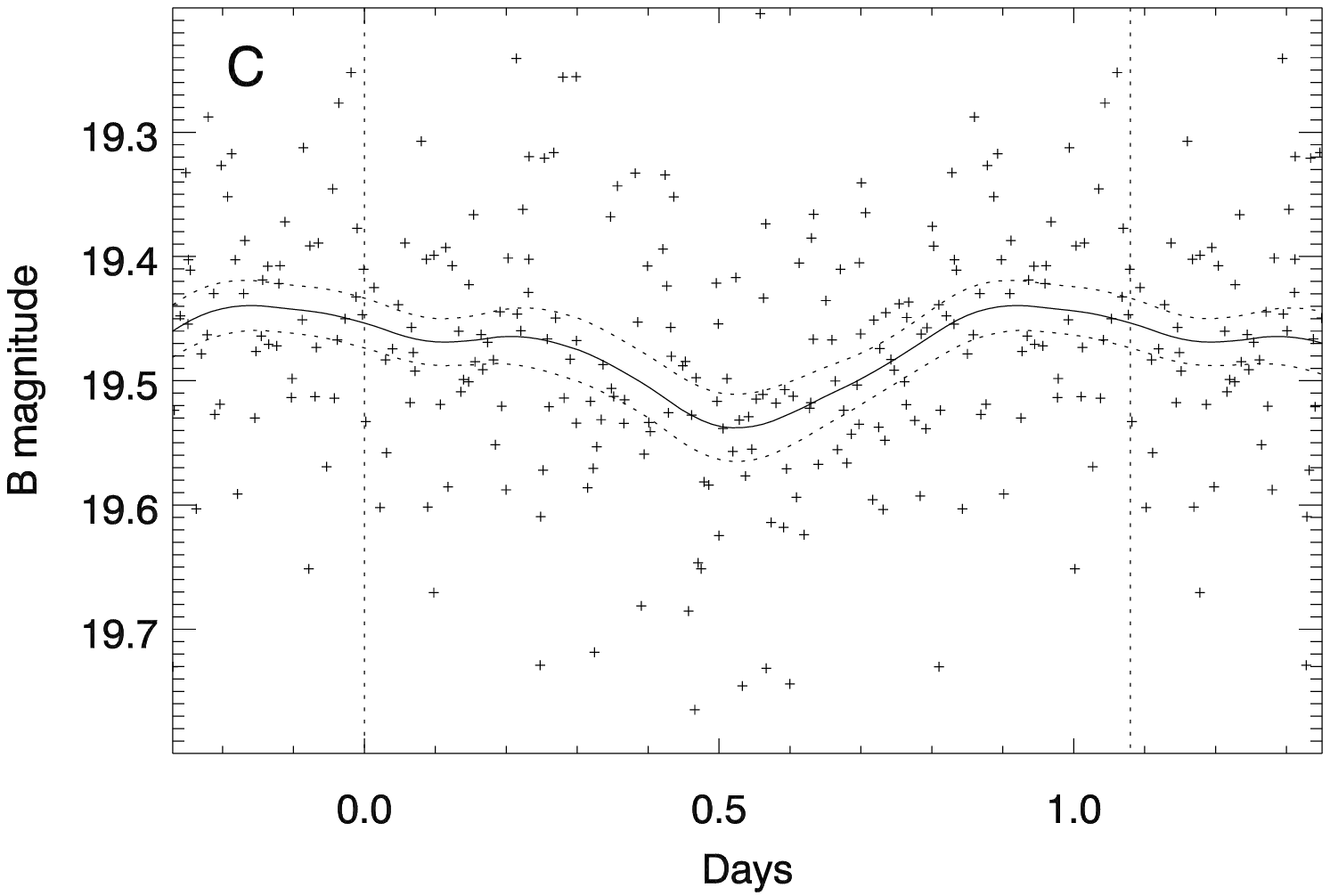}
\caption[Swift periodogram]{
	\label{swift_periodogram}
	\label{lastfig}			
	(A) Periodogram of $Swift$ data calculated using the FASPER
algorithm of \citet{numrecipes}.  The significance levels are calculated
via a Monte Carlo approach as suggested in \citet{numrecipes}.
(B)$Swift$ measurements of Eris
 phased to a period of 15.0 days.  Continuously overplotted is a running
mean of the nearest 20 phased data points, 
along with the 1$\sigma$ estimated uncertainty in that
mean.
(C)$Swift$ measurements of Eris
 phased to a period of 1.08 days.  Continuously overplotted is a running
mean of the nearest 20 phased data points, 
along with the 1$\sigma$ estimated uncertainty in that
mean.
We strongly
discount the calculated significance level of the 15 day period as 
our dataset covered only 28 days.
	}
\end{center}
\end{figure}

The Lomb-Scargle
periodogram \citep{numrecipes} for these data is shown in
Fig.~\ref{swift_periodogram}A.  The significance levels are calculated
following the technique suggested by \citet{numrecipes}, which is to
shuffle the data, randomly reassigning observed magnitudes to
observation times, and recalculate the periodogram.  This operation is
performed repeatedly and the peak power in each shuffled periodogram
is recorded.  The significance levels are determined from $1-f$, where
$f$ is the fraction of the samples for which the highest peak is
greater than the power level in question.  Two peaks (at $\sim$1.1 and
$\sim$15 days) are at suggestive levels of significance.  The data are
shown
phased to these periods in Fig.~\ref{swift_periodogram}B and
\ref{swift_periodogram}C.  

Periodograms are  powerful tools, but
have serious limitations.  These data and the $\sim$15 day peak
are a case study in some of the hazards of  periodograms if incorrectly
interpreted.
One should be extremely wary of strong peaks at
periods longer than $\sim$1/3-1/2 of the total time period covered. 
The phased data of Fig.~\ref{swift_periodogram}B show significant gaps
in data coverage, which is an additional warning sign in periodogram
analysis that results may not be trustworthy.
From Fig.~\ref{swift_periodogram}A it would be very tempting
to conclude that we detected a $\sim$15 day period in Eris, but the
data incompletely
span only approximately twice this length of time.  
Examination of the statistics of the variability of the daily mean 
does not support the detection of a 15 day
period.  Although the human eye is drawn to the three consecutive
daily mean
magnitudes in Fig.~\ref{swift_fluxes} that lie $>1\sigma$ fainter than
the overall mean of the dataset (JD 2454100-2454102), these appear
to be a statistical fluke.
A simple Monte Carlo simulation reveals that in a Gaussian noise dataset
of 23 points (the number of daily means in Fig.~\ref{swift_fluxes})
about 36\% of the time there will be three consecutive data points that
all
sit 1-3$\sigma$ above the mean or all sit 1-3$\sigma$ below the mean.
Due to these issues we strongly discount the significance of the
peak at $\sim$15 days.  

More interesting is the peak at $\sim$1.1 days, which has
a significance level of 97$\%$.  There
are no obvious issues in the phased data of Fig.~\ref{swift_periodogram}C.
To probe the validity of this possible $\sim$1.1 day period
 we ran a variety of tests on our data.
We split the dataset in half and found a similar result from each
half, although with
the expected lower confidence level due to fewer data.  We ran a sequence
of tests in which we randomly selected half of the data points to analyze
and again found similar results.  We searched the data for any possible
correlations, e.g.\ flux of Eris with position on the detector, but identified
none that could explain the signal seen in Fig.~\ref{swift_periodogram}C
or that could not be ruled out by examination of the ensemble of 
comparison stars.  We examined the Lomb-Scargle periodogram for each of the
26 comparison stars.  As expected, given that we had eliminated potential
comparison stars with obvious periodicities, none of the 26 comparison
stars displayed peaks in the periodogram above a significance level of
90\%.

\begin{figure}[p!]
\begin{center}
\includegraphics[width=4in]{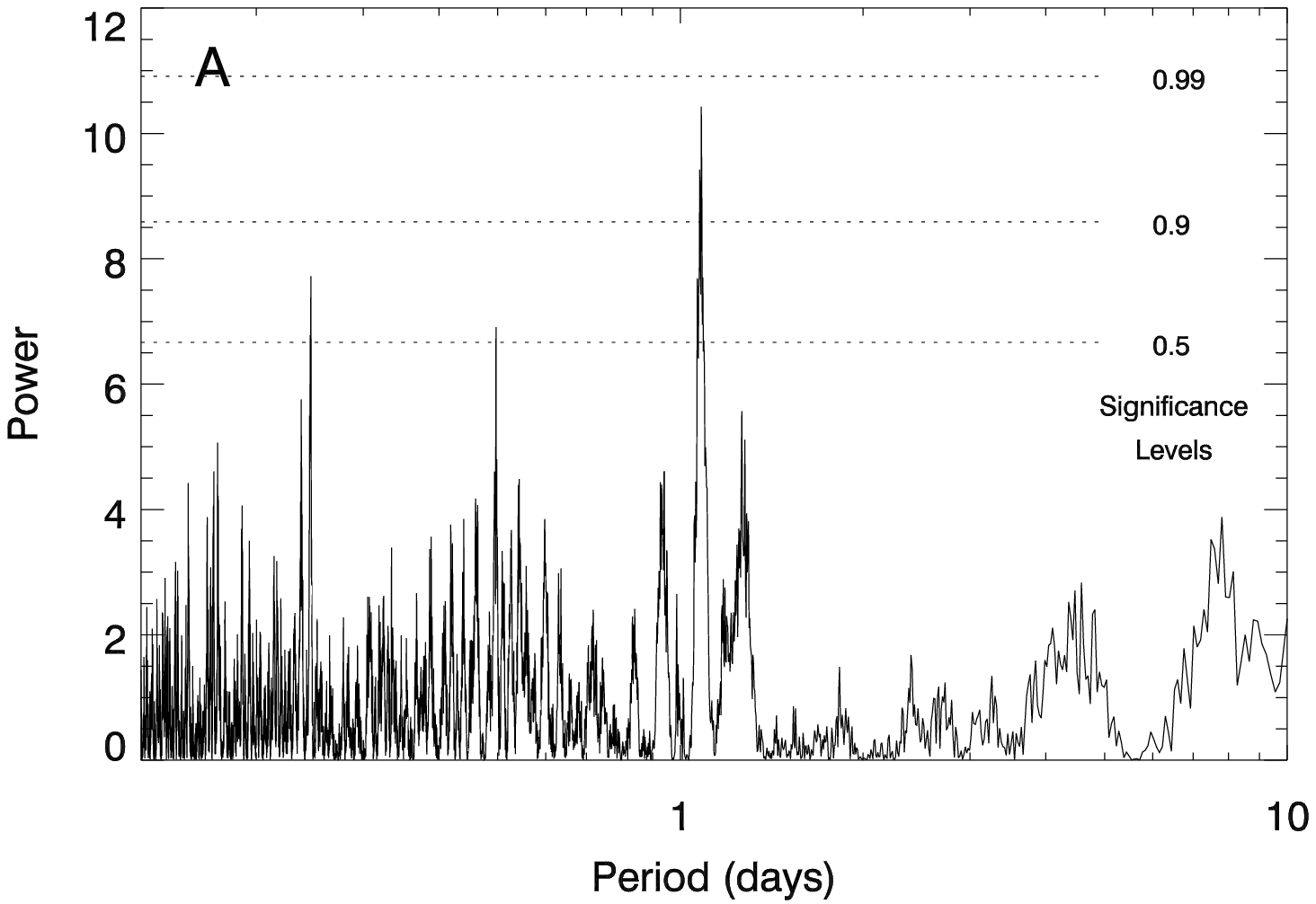}
\includegraphics[width=4in]{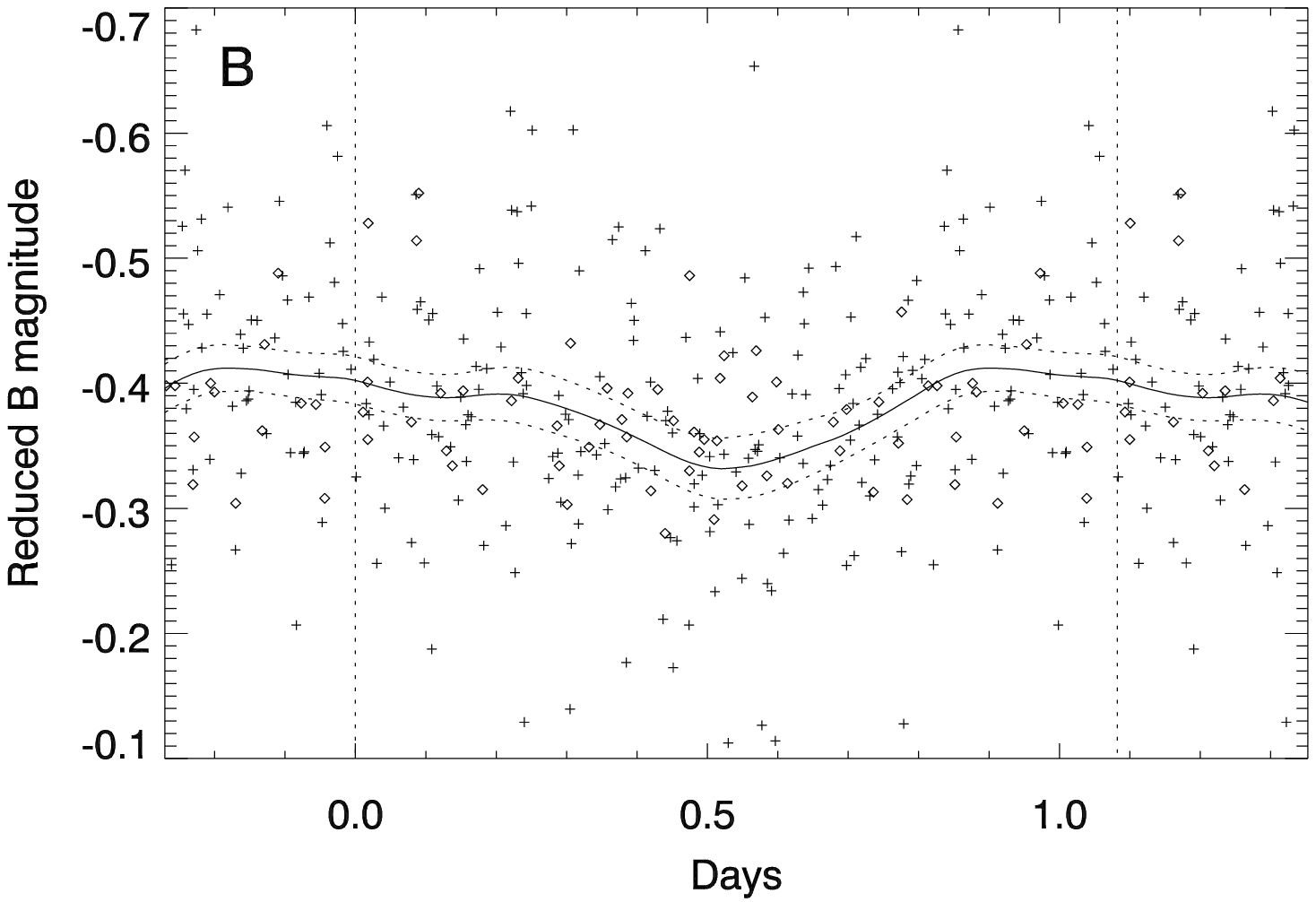}
\caption[Swift combined data]{
	\label{combinedfig}
	\label{lastfig}			
	(A) Periodogram of the combined $Swift$ and \citet{2007AJ....133...26R}
dataset
 calculated using the FASPER
algorithm of \citet{numrecipes}.  The significance levels are calculated
via a Monte Carlo approach as suggested in \citet{numrecipes}.
(C) Combined dataset
 phased to a period of 1.08 days.  $Swift$ data are shown as plus signs (+),
while \citet{2007AJ....133...26R} data are shown as diamonds ($\diamond$).
Continuously overplotted is a running
mean of the nearest 20 phased data points, 
along with the 1$\sigma$ estimated uncertainty in that
mean.  The addition of the \citet{2007AJ....133...26R} data slightly
improve the statistical significance of the recovered periodicity.

	}
\end{center}
\end{figure}

To further test the validity of the possible $\sim$1.1 day period we
combined the $Swift$ dataset with the B-band measurements of 
\citet{2007AJ....133...26R}, which is the one other published dataset
with a significant number of high-quality B-band measurements.  
After scaling the $Swift$ measurements to the reduced magnitude system
of \citet{2007AJ....133...26R} the periodogram of the combined dataset
is shown in Fig.~\ref{combinedfig}A.  The peak of the periodogram
remains at approximately the same period (1.08~days) as with the
$Swift$ dataset alone, although the significance level of the peak
increases slightly to nearly 99\%.  The combined data phased
to this 1.08~day period are shown in Fig.~\ref{combinedfig}B.  The
results from the
combined dataset look very similar to the results from the $Swift$
dataset alone.

We can find no 
reason to discount the validity of the
 $\sim$1.1 day period.  
  Using the 50$\%$ confidence levels of the 
periodogram as a guide
for estimating the uncertainty in the period, we report that
Eris appears to be rotating once every 1.08$\pm$0.02 days with a 
peak-to-valley amplitude of nearly 0.1 mag.  At the modest level of 
confidence available from these data the periodic signal appears less
like a sinusoid and more like what would be expected from a large
dark patch on the partially hidden hemisphere of an otherwise homogeneous 
body.  At rotational phases where the patch is hidden from the observer
the light curve is constant.  A dip in the light curve is observed
during the rotational phases that the dark patch is visible to the observer.
Additional observations will be required to confirm this 
result and more precisely determine the
rotation period of Eris.  

Of the previously published datasets the long-term sequence of
\citet{2007AJ....133...26R} and the several nights of precision
photometry of \citet{2007AJ....134..787S} are the most likely to have
been able to identify the apparent periodicity seen in these {\it Swift} data.

\citet{2007AJ....134..787S} did precision $R$-band photometry of Eris
over several hours on each of three contiguous nights in October 2005
and three contiguous nights in December 2005.  A careful evaluation of
the approximately daily sampling of their data reveals that a periodic
signal such as that suggested by the {\it Swift} data in
Fig.~\ref{swift_periodogram}C could have been missed.  The signal in
Fig.~\ref{swift_periodogram}C appears nearly constant over $\sim$50\%
of the rotation period, while phased to a period of 1.08 days the
\citet{2007AJ....134..787S} data cover less than half of this rotation
period.  

The magnitude of variation in the light curve of Eris is
likely a function of wavelength.  Most of the surface of Eris must be
uniformly bright to achieve the high $V$-band geometric albedo
\citep[86\%$\pm$7\%;][]{2006ApJ...643L..61B}.  Any darker patches are
likely to be photochemically processed hydrocarbons, which will have a
red color.  At longer wavelengths (e.g.\ $R$-band and $I$-band) the
albedo of these red patches will be closer to the albedo of the
uniformly bright dominant surface material.  At shorter wavelengths
(e.g.\ the $B$-band used here by $Swift$) there will be greater
contrast between these red patches and the rest of the surface of
Eris.  This is directly analogous to what is seen on Pluto, where the
light curve at $B$-band has an amplitude of $\sim$0.1~mag greater than
the amplitude at $R$-band \citep{2003Icar..162..171B}.  Thus, the
light curve of Eris is likely to be less pronounced at longer
wavelengths (e.g.\ R-band) than the shorter wavelength B-band
observed with $Swift$.

\citet{2007AJ....133...26R} reported observations in several filters,
taken once per night over many months.  The analysis of the
\citet{2007AJ....133...26R} data is thus dependent upon the
simultaneous fitting of color terms and phase function coefficients.
We experimented with inserting fake signals into the
\citet{2007AJ....133...26R} data with a range of periods and
amplitudes consistent with the periodicity detected in the {\it Swift}
data.  We then used the same analysis tools as before to search for
periodicity.  In many cases the inserted periodicity is recovered,
however whether the fake signal is detected and the period at which it
is detected is strongly dependent on the exact amplitude and period of
the fake signal.  This is primarily because the test periods
($\sim$1.08 days) are near the sampling interval ($\sim$1 day) of the
\citet{2007AJ....133...26R} data.  In some example cases varying the
inserted signal's period by only 0.005 days made the difference
between a strong detection and a non-detection.  However, as
described above, the combination of the B-band measurements of
\citet{2007AJ....133...26R} with the $Swift$ measurements reported
here improves the significance of the retrieved periodicity somewhat
from that of the $Swift$ data alone.

A coherent picture of Eris is emerging.  The surface is primarily
covered in bright methane frost, much like the brightest patches of
Pluto's surface.  However, our results suggest that the surface of 
Eris is not perfectly homogeneous.
Under thermal equilibrium the vapor pressure of
methane on Eris is negligible at its current near-perihelion distance
of 97.5 AU.  If the entire surface of Eris is uniformly
covered in very high albedo methane frost, even at aphelion (38.2 AU)
the equilibrium surface temperature of Eris is not warm enough to
generate significant methane evaporation from the surface. This presents
a problem as methane frost
in the outer solar system is expected to redden and darken due to
photochemical processing, but Eris appears bright and shows no sign of
redness.  This suggests the methane frost deposits on its surface must
be fresh and a replenishment mechanism is required.  Our detection of
variability is consistent with regional darker areas on the surface.  At
aphelion the widespread high albedo regions will not warm enough to 
sublimate methane into the atmosphere, however the small darker areas
will be heated by the increased insolation to kickoff feedback effects
that lead to dramatic global surface change, generating a temporary 
atmosphere and replenishing the methane
surface each Eris year.


\section{Acknowledgements}

We thank David Rabinowitz and an anonymous referee.

\label{lastpage}


\bibliography{I10479_bib.bib}

 \bibliographystyle{I10479_elsart-harv}


\clearpage	


\clearpage

{
\renewcommand{\baselinestretch}{1}
\small\normalsize
\begin{center}
\begin{longtable}{cccc}
\caption[Swift measurements]{Table 1: Swift measurements. \label{swiftmaintable}} \\
\hline
\hline
Julian Date & Exp.\ time (s) & B Magnitude & Phase angle \\
\hline
2454088.576 &  390.5 & 19.40$\pm$0.09 & 0$\fdg$535 \\
2454088.643 &  390.5 & 19.37$\pm$0.09 & 0$\fdg$536 \\
2454088.710 &  391.0 & 19.46$\pm$0.09 & 0$\fdg$536 \\
2454093.585 &  450.4 & 19.53$\pm$0.10 & 0$\fdg$553 \\
2454093.652 &  450.9 & 19.18$\pm$0.08 & 0$\fdg$553 \\
2454093.719 &  450.8 & 19.43$\pm$0.09 & 0$\fdg$553 \\
2454093.786 &  450.9 & 19.51$\pm$0.10 & 0$\fdg$553 \\
2454093.853 &  450.5 & 19.28$\pm$0.08 & 0$\fdg$554 \\
2454093.920 &  450.4 & 19.56$\pm$0.10 & 0$\fdg$554 \\
2454093.987 &  450.9 & 19.67$\pm$0.11 & 0$\fdg$554 \\
2454094.054 &  451.0 & 19.46$\pm$0.10 & 0$\fdg$554 \\
2454094.121 &  451.0 & 19.40$\pm$0.09 & 0$\fdg$554 \\
2454094.188 &  451.0 & 19.26$\pm$0.08 & 0$\fdg$555 \\
2454094.255 &  451.5 & 19.53$\pm$0.10 & 0$\fdg$555 \\
2454094.322 &  450.8 & 19.48$\pm$0.10 & 0$\fdg$555 \\
2454094.389 &  450.9 & 19.62$\pm$0.10 & 0$\fdg$555 \\
2454094.456 &  450.4 & 19.73$\pm$0.11 & 0$\fdg$555 \\
2454094.522 &  450.9 & 19.37$\pm$0.09 & 0$\fdg$556 \\
2454094.590 &  451.0 & 19.34$\pm$0.09 & 0$\fdg$556 \\
2454094.656 &  450.9 & 19.44$\pm$0.09 & 0$\fdg$556 \\
2454094.724 &  450.9 & 19.41$\pm$0.09 & 0$\fdg$556 \\
2454094.790 &  450.4 & 19.59$\pm$0.10 & 0$\fdg$557 \\
2454094.857 &  450.4 & 19.37$\pm$0.09 & 0$\fdg$557 \\
2454094.924 &  450.4 & 19.35$\pm$0.09 & 0$\fdg$557 \\
2454094.991 &  449.9 & 19.60$\pm$0.11 & 0$\fdg$557 \\
2454095.058 &  449.4 & 19.60$\pm$0.10 & 0$\fdg$557 \\
2454095.125 &  450.8 & 19.48$\pm$0.09 & 0$\fdg$557 \\
2454095.192 &  450.9 & 19.36$\pm$0.09 & 0$\fdg$558 \\
2454095.259 &  450.4 & 19.48$\pm$0.10 & 0$\fdg$558 \\
2454095.326 &  450.5 & 19.34$\pm$0.09 & 0$\fdg$558 \\
2454095.393 &  450.5 & 19.33$\pm$0.09 & 0$\fdg$558 \\
2454095.460 &  450.5 & 19.86$\pm$0.13 & 0$\fdg$558 \\
2454095.527 &  450.4 & 19.20$\pm$0.08 & 0$\fdg$559 \\
2454095.597 &  922.9 & 19.52$\pm$0.07 & 0$\fdg$559 \\
2454095.664 &  863.9 & 19.50$\pm$0.07 & 0$\fdg$559 \\
2454095.731 &  921.9 & 19.50$\pm$0.07 & 0$\fdg$559 \\
2454095.797 &  923.4 & 19.33$\pm$0.06 & 0$\fdg$559 \\
2454095.862 &  449.0 & 19.32$\pm$0.08 & 0$\fdg$559 \\
2454095.929 &  450.0 & 19.42$\pm$0.09 & 0$\fdg$560 \\
2454095.996 &  450.1 & 19.57$\pm$0.10 & 0$\fdg$560 \\
2454096.063 &  450.5 & 19.43$\pm$0.09 & 0$\fdg$560 \\
2454096.129 &  450.5 & 19.31$\pm$0.08 & 0$\fdg$560 \\
2454096.197 &  450.5 & 19.42$\pm$0.09 & 0$\fdg$560 \\
2454096.263 &  451.0 & 19.24$\pm$0.08 & 0$\fdg$561 \\
2454096.331 &  450.5 & 19.51$\pm$0.10 & 0$\fdg$561 \\
2454096.397 &  450.4 & 19.51$\pm$0.10 & 0$\fdg$561 \\
2454096.546 & 1337.1 & 19.52$\pm$0.06 & 0$\fdg$561 \\
2454096.611 &  982.2 & 19.51$\pm$0.06 & 0$\fdg$561 \\
2454096.679 & 1159.4 & 19.39$\pm$0.05 & 0$\fdg$562 \\
2454096.747 & 1336.2 & 19.41$\pm$0.05 & 0$\fdg$562 \\
2454096.814 & 1336.2 & 19.45$\pm$0.05 & 0$\fdg$562 \\
2454096.881 & 1336.6 & 19.40$\pm$0.05 & 0$\fdg$562 \\
2454099.224 & 1311.0 & 19.39$\pm$0.05 & 0$\fdg$568 \\
2454099.291 & 1370.8 & 19.53$\pm$0.06 & 0$\fdg$568 \\
2454099.358 & 1312.3 & 19.48$\pm$0.05 & 0$\fdg$568 \\
2454099.425 & 1371.5 & 19.51$\pm$0.05 & 0$\fdg$568 \\
2454099.492 & 1372.5 & 19.40$\pm$0.05 & 0$\fdg$568 \\
2454099.764 &  574.3 & 19.65$\pm$0.09 & 0$\fdg$569 \\
2454099.831 &  632.7 & 19.53$\pm$0.08 & 0$\fdg$569 \\
2454099.898 &  632.8 & 19.59$\pm$0.08 & 0$\fdg$569 \\
2454099.965 &  632.7 & 19.52$\pm$0.08 & 0$\fdg$569 \\
2454100.032 &  632.7 & 19.48$\pm$0.08 & 0$\fdg$569 \\
2454100.099 &  633.3 & 19.44$\pm$0.08 & 0$\fdg$569 \\
2454100.166 &  632.6 & 19.52$\pm$0.08 & 0$\fdg$570 \\
2454100.233 &  632.7 & 19.41$\pm$0.07 & 0$\fdg$570 \\
2454100.300 &  632.7 & 19.51$\pm$0.08 & 0$\fdg$570 \\
2454100.366 &  632.1 & 19.45$\pm$0.07 & 0$\fdg$570 \\
2454100.434 &  632.0 & 19.52$\pm$0.08 & 0$\fdg$570 \\
2454100.768 &  634.6 & 19.41$\pm$0.07 & 0$\fdg$571 \\
2454100.835 &  634.5 & 19.76$\pm$0.09 & 0$\fdg$571 \\
2454100.902 &  693.0 & 19.75$\pm$0.09 & 0$\fdg$571 \\
2454100.969 &  634.6 & 19.74$\pm$0.10 & 0$\fdg$571 \\
2454101.036 &  693.0 & 19.56$\pm$0.08 & 0$\fdg$571 \\
2454101.103 &  634.5 & 19.55$\pm$0.08 & 0$\fdg$571 \\
2454101.170 &  634.5 & 19.38$\pm$0.07 & 0$\fdg$572 \\
2454101.237 &  693.5 & 19.43$\pm$0.07 & 0$\fdg$572 \\
2454101.304 &  634.7 & 19.46$\pm$0.08 & 0$\fdg$572 \\
2454101.371 &  693.1 & 19.65$\pm$0.08 & 0$\fdg$572 \\
2454101.438 &  693.1 & 19.43$\pm$0.07 & 0$\fdg$572 \\
2454101.772 &  634.2 & 19.57$\pm$0.08 & 0$\fdg$573 \\
2454101.840 &  693.5 & 19.68$\pm$0.08 & 0$\fdg$573 \\
2454101.906 &  693.7 & 19.69$\pm$0.08 & 0$\fdg$573 \\
2454101.973 &  693.2 & 19.42$\pm$0.07 & 0$\fdg$573 \\
2454102.040 &  693.5 & 19.62$\pm$0.08 & 0$\fdg$573 \\
2454102.107 &  693.7 & 19.85$\pm$0.10 & 0$\fdg$573 \\
2454102.174 &  693.0 & 19.54$\pm$0.08 & 0$\fdg$573 \\
2454102.241 &  693.2 & 19.54$\pm$0.08 & 0$\fdg$573 \\
2454102.308 &  693.4 & 19.46$\pm$0.07 & 0$\fdg$574 \\
2454102.375 &  693.6 & 19.53$\pm$0.07 & 0$\fdg$574 \\
2454102.442 &  693.5 & 19.45$\pm$0.07 & 0$\fdg$574 \\
2454102.777 &  575.2 & 19.73$\pm$0.10 & 0$\fdg$574 \\
2454102.844 &  634.1 & 19.59$\pm$0.08 & 0$\fdg$574 \\
2454102.911 &  574.6 & 19.33$\pm$0.07 & 0$\fdg$574 \\
2454102.978 &  634.2 & 19.49$\pm$0.08 & 0$\fdg$575 \\
2454103.246 &  575.2 & 19.60$\pm$0.09 & 0$\fdg$575 \\
2454103.313 &  634.2 & 19.59$\pm$0.09 & 0$\fdg$575 \\
2454103.380 &  634.2 & 19.48$\pm$0.08 & 0$\fdg$575 \\
2454103.571 &  664.4 & 19.47$\pm$0.07 & 0$\fdg$575 \\
2454103.639 &  900.2 & 19.48$\pm$0.06 & 0$\fdg$576 \\
2454103.708 & 1195.2 & 19.40$\pm$0.05 & 0$\fdg$576 \\
2454103.776 & 1431.9 & 19.49$\pm$0.05 & 0$\fdg$576 \\
2454103.842 & 1196.2 & 19.32$\pm$0.05 & 0$\fdg$576 \\
2454103.909 & 1254.8 & 19.47$\pm$0.06 & 0$\fdg$576 \\
2454105.121 &  403.3 & 19.46$\pm$0.09 & 0$\fdg$577 \\
2454105.188 &  439.0 & 19.45$\pm$0.09 & 0$\fdg$578 \\
2454105.255 &  416.2 & 19.37$\pm$0.09 & 0$\fdg$578 \\
2454105.322 &  452.0 & 19.47$\pm$0.09 & 0$\fdg$578 \\
2454105.389 &  428.7 & 19.46$\pm$0.09 & 0$\fdg$578 \\
2454106.582 &  545.9 & 19.52$\pm$0.08 & 0$\fdg$579 \\
2454106.648 &  486.1 & 19.33$\pm$0.08 & 0$\fdg$579 \\
2454106.715 &  486.6 & 19.47$\pm$0.09 & 0$\fdg$579 \\
2454106.782 &  487.5 & 19.47$\pm$0.08 & 0$\fdg$579 \\
2454106.849 &  487.1 & 19.41$\pm$0.08 & 0$\fdg$579 \\
2454106.916 &  487.1 & 19.46$\pm$0.08 & 0$\fdg$579 \\
2454106.983 &  487.1 & 19.46$\pm$0.09 & 0$\fdg$579 \\
2454107.049 &  486.5 & 19.59$\pm$0.09 & 0$\fdg$579 \\
2454107.117 &  487.0 & 19.32$\pm$0.08 & 0$\fdg$580 \\
2454107.183 &  487.0 & 19.53$\pm$0.09 & 0$\fdg$580 \\
2454107.251 &  486.5 & 19.53$\pm$0.09 & 0$\fdg$580 \\
2454107.317 &  486.7 & 19.50$\pm$0.09 & 0$\fdg$580 \\
2454107.585 &  487.1 & 19.45$\pm$0.08 & 0$\fdg$580 \\
2454107.652 &  487.2 & 19.39$\pm$0.08 & 0$\fdg$580 \\
2454107.719 &  487.1 & 19.53$\pm$0.09 & 0$\fdg$580 \\
2454107.786 &  487.1 & 19.42$\pm$0.08 & 0$\fdg$580 \\
2454107.853 &  487.2 & 19.39$\pm$0.08 & 0$\fdg$580 \\
2454107.919 &  487.2 & 19.38$\pm$0.08 & 0$\fdg$580 \\
2454107.987 &  487.0 & 19.39$\pm$0.08 & 0$\fdg$580 \\
2454108.053 &  487.0 & 19.41$\pm$0.08 & 0$\fdg$580 \\
2454108.121 &  487.7 & 19.44$\pm$0.09 & 0$\fdg$580 \\
2454108.187 &  487.1 & 19.47$\pm$0.08 & 0$\fdg$580 \\
2454108.254 &  487.6 & 19.72$\pm$0.10 & 0$\fdg$580 \\
2454108.522 &  487.0 & 19.51$\pm$0.09 & 0$\fdg$580 \\
2454108.589 &  486.6 & 19.47$\pm$0.09 & 0$\fdg$581 \\
2454108.656 &  486.1 & 19.47$\pm$0.08 & 0$\fdg$581 \\
2454108.723 &  486.6 & 19.46$\pm$0.09 & 0$\fdg$581 \\
2454108.789 &  487.5 & 19.29$\pm$0.07 & 0$\fdg$581 \\
2454108.857 &  487.0 & 19.48$\pm$0.09 & 0$\fdg$581 \\
2454108.924 &  487.1 & 19.31$\pm$0.08 & 0$\fdg$581 \\
2454108.991 &  487.0 & 19.25$\pm$0.07 & 0$\fdg$581 \\
2454109.058 &  486.6 & 19.44$\pm$0.08 & 0$\fdg$581 \\
2454109.124 &  487.6 & 19.39$\pm$0.08 & 0$\fdg$581 \\
2454109.192 &  487.0 & 19.48$\pm$0.09 & 0$\fdg$581 \\
2454109.258 &  487.0 & 19.61$\pm$0.09 & 0$\fdg$581 \\
2454109.521 & 1089.7 & 19.50$\pm$0.07 & 0$\fdg$581 \\
2454109.590 & 1355.8 & 19.52$\pm$0.06 & 0$\fdg$581 \\
2454109.660 & 1571.7 & 19.44$\pm$0.05 & 0$\fdg$581 \\
2454109.728 & 1748.0 & 19.45$\pm$0.05 & 0$\fdg$581 \\
2454109.795 & 1749.5 & 19.46$\pm$0.05 & 0$\fdg$581 \\
2454109.853 &  617.8 & 19.60$\pm$0.10 & 0$\fdg$581 \\
2454109.921 &  671.8 & 19.39$\pm$0.08 & 0$\fdg$581 \\
2454109.988 &  553.6 & 19.50$\pm$0.09 & 0$\fdg$581 \\
2454110.063 & 1749.6 & 19.45$\pm$0.05 & 0$\fdg$581 \\
2454110.129 & 1750.0 & 19.47$\pm$0.05 & 0$\fdg$581 \\
2454110.197 & 1749.4 & 19.52$\pm$0.05 & 0$\fdg$581 \\
2454110.263 & 1749.3 & 19.47$\pm$0.05 & 0$\fdg$581 \\
2454110.321 &  488.2 & 19.43$\pm$0.10 & 0$\fdg$581 \\
2454110.344 &  384.0 & 19.32$\pm$0.09 & 0$\fdg$581 \\
2454110.388 &  665.2 & 19.53$\pm$0.10 & 0$\fdg$581 \\
2454110.457 &  901.4 & 19.52$\pm$0.08 & 0$\fdg$581 \\
2454110.526 & 1043.6 & 19.35$\pm$0.06 & 0$\fdg$581 \\
2454110.595 & 1515.9 & 19.54$\pm$0.06 & 0$\fdg$581 \\
2454110.663 & 1575.4 & 19.61$\pm$0.06 & 0$\fdg$581 \\
2454110.730 & 1575.7 & 19.57$\pm$0.06 & 0$\fdg$581 \\
2454110.796 & 1575.9 & 19.36$\pm$0.05 & 0$\fdg$581 \\
2454111.598 & 1452.3 & 19.53$\pm$0.06 & 0$\fdg$581 \\
2454111.666 & 1452.3 & 19.42$\pm$0.06 & 0$\fdg$582 \\
2454111.732 & 1452.4 & 19.43$\pm$0.05 & 0$\fdg$582 \\
2454111.799 & 1451.9 & 19.52$\pm$0.06 & 0$\fdg$582 \\
2454111.867 & 1363.2 & 19.53$\pm$0.06 & 0$\fdg$582 \\
2454111.934 & 1364.2 & 19.52$\pm$0.06 & 0$\fdg$582 \\
2454112.001 & 1364.1 & 19.45$\pm$0.06 & 0$\fdg$582 \\
2454112.068 & 1364.2 & 19.40$\pm$0.06 & 0$\fdg$582 \\
2454112.131 &  753.5 & 19.41$\pm$0.07 & 0$\fdg$582 \\
2454112.397 &  546.4 & 19.50$\pm$0.10 & 0$\fdg$582 \\
2454112.465 &  841.1 & 19.45$\pm$0.08 & 0$\fdg$582 \\
2454112.530 &  389.5 & 19.26$\pm$0.10 & 0$\fdg$582 \\
2454112.597 &  449.0 & 19.37$\pm$0.10 & 0$\fdg$581 \\
2454112.671 &  980.2 & 19.39$\pm$0.06 & 0$\fdg$581 \\
2454112.735 &  980.2 & 19.58$\pm$0.07 & 0$\fdg$581 \\
2454112.802 &  929.6 & 19.51$\pm$0.07 & 0$\fdg$581 \\
2454112.870 &  921.2 & 19.62$\pm$0.08 & 0$\fdg$581 \\
2454112.936 &  921.4 & 19.54$\pm$0.07 & 0$\fdg$581 \\
2454113.004 &  921.2 & 19.44$\pm$0.07 & 0$\fdg$581 \\
2454113.070 &  920.8 & 19.45$\pm$0.07 & 0$\fdg$581 \\
2454113.137 &  921.3 & 19.35$\pm$0.06 & 0$\fdg$581 \\
2454113.206 &  597.1 & 19.47$\pm$0.08 & 0$\fdg$581 \\
2454113.287 &  613.2 & 19.51$\pm$0.08 & 0$\fdg$581 \\
2454113.401 &  512.4 & 19.49$\pm$0.10 & 0$\fdg$581 \\
2454113.470 &  808.0 & 19.50$\pm$0.08 & 0$\fdg$581 \\
2454113.681 & 1438.5 & 19.51$\pm$0.06 & 0$\fdg$581 \\
2454113.756 &  611.5 & 19.42$\pm$0.08 & 0$\fdg$581 \\
2454113.809 & 1599.0 & 19.58$\pm$0.06 & 0$\fdg$581 \\
2454113.876 & 1599.1 & 19.56$\pm$0.06 & 0$\fdg$581 \\
2454113.943 & 1598.8 & 19.41$\pm$0.05 & 0$\fdg$581 \\
2454114.010 & 1598.5 & 19.57$\pm$0.06 & 0$\fdg$581 \\
2454114.077 & 1598.0 & 19.49$\pm$0.05 & 0$\fdg$581 \\
2454114.140 & 1008.6 & 19.73$\pm$0.09 & 0$\fdg$581 \\
2454114.679 & 1639.6 & 19.45$\pm$0.05 & 0$\fdg$581 \\
2454114.746 & 1577.3 & 19.49$\pm$0.05 & 0$\fdg$581 \\
2454114.813 & 1576.9 & 19.54$\pm$0.06 & 0$\fdg$581 \\
2454114.880 & 1577.4 & 19.65$\pm$0.06 & 0$\fdg$581 \\
2454114.947 & 1577.3 & 19.58$\pm$0.06 & 0$\fdg$581 \\
2454115.014 & 1577.3 & 19.51$\pm$0.05 & 0$\fdg$581 \\
2454115.081 & 1576.9 & 19.41$\pm$0.05 & 0$\fdg$581 \\
2454115.141 &  565.6 & 19.60$\pm$0.11 & 0$\fdg$581 \\
2454115.683 & 1629.4 & 19.52$\pm$0.05 & 0$\fdg$580 \\
2454115.750 & 1630.0 & 19.52$\pm$0.06 & 0$\fdg$580 \\
2454115.818 & 1629.1 & 19.55$\pm$0.06 & 0$\fdg$580 \\
2454115.884 & 1591.1 & 19.56$\pm$0.06 & 0$\fdg$580 \\
2454115.951 & 1591.9 & 19.53$\pm$0.06 & 0$\fdg$580 \\
2454116.018 & 1595.2 & 19.53$\pm$0.06 & 0$\fdg$580 \\
2454116.085 & 1595.0 & 19.57$\pm$0.06 & 0$\fdg$580 \\
2454116.154 &  560.4 & 19.50$\pm$0.08 & 0$\fdg$580 \\
2454116.688 & 1594.7 & 19.59$\pm$0.06 & 0$\fdg$580 \\
2454116.755 & 1594.2 & 19.55$\pm$0.06 & 0$\fdg$580 \\
2454116.822 & 1594.2 & 19.57$\pm$0.06 & 0$\fdg$580 \\
2454116.889 & 1594.0 & 19.52$\pm$0.05 & 0$\fdg$580 \\
2454116.955 & 1594.6 & 19.45$\pm$0.05 & 0$\fdg$580 \\
2454117.023 & 1593.2 & 19.48$\pm$0.05 & 0$\fdg$580 \\
2454117.089 & 1594.6 & 19.56$\pm$0.06 & 0$\fdg$579 \\
\hline
\end{longtable}
\end{center}
}

\end{document}